\documentstyle[prl,aps,multicol,epsfig]{revtex}
\newcommand{\beq}{\begin{equation}}
\newcommand{\eeq}{\end{equation}}

\begin{document}

\title{C-Axis Tunneling Spectra in High-T$_c$ Superconductors
in the Presence of a \\d Charge-Density Wave}
\author{{A. Greco}$^{a,b}$ and {R. Zeyher}$^a$ } 
\address{$^a$Max-Planck-Institut\  f\"ur\
Festk\"orperforschung,\\ Heisenbergstr.1, 70569 Stuttgart, Germany \\
$^b$Permanent address: Departamento de F\'{\i}sica, Facultad de
Ciencias Exactas e Ingenier\'{\i}a and \\
IFIR(UNR-CONICET), Av. Pellegrini 250, 
2000-Rosario, Argentina}
\date{\today}
\maketitle
\begin{abstract} 
The optimally doped and underdoped region of the $t-J$ model at large N
(N is the number of spin components) is governed by the competition of
d-wave superconductivity (SC) and a d Charge-Density Wave (d-CDW).
The partial destruction of the Fermi surface by the d-CDW and the resulting 
density of states are discussed. Furthermore, c-axis
conductances for incoherent and coherent tunneling are calculated, 
considering both
an isotropic and an anisotropic in-plane momentum dependence of the
hopping matrix element between the planes. The influence of self-energy
effects on the conductances is also considered using a model where the
electrons interact with a dispersionless, low-lying branch of bosons.
\\ \indent
We show that available tunneling spectra from break-junctions are best
explained by assuming that they result from incoherent tunneling
with a strongly anisotropic hopping matrix element of the form suggested by
band structure calculations. The conductance spectra are then characterized
by one single peak which evolves continuously from the superconducting
to the d-CDW state with decreasing doping. The intrinsic c-axis
tunneling spectra are, on the other hand, best explained 
by coherent tunneling. Calculated spectra show at low temperatures two peaks 
due to SC and d-CDW. With increasing temperature the BCS-like peak
moves to zero voltage and vanishes at T$_c$, exactly as in experiment.
Our results thus can explain why break junction and intrinsic 
tunneling spectra are different from each other. Moreover, they
support a scenario of two competing order parameters
in the underdoped region of high-T$_c$ superconductors.
\end{abstract}
\pacs{74.72.-h,74.50.+r,71.10.Hf}

\begin{multicols}{2}


\section{Introduction}

One important topic in high-T$_c$ superconductors is the question of
how many order parameters are needed for a
proper description of the optimally doped and underdoped cases.
One scenario assumes that only the superconducting order parammeter
is relevant. The decrease in the transition temperature T$_c$ is
then caused by fluctuations of its phase and the pseudogap
is locally just the superconducting gap. A second scenario 
assumes that the physics in the underdoped and optimally doped region
is mainly determined by the competition of the superconducting
order parameter with a second one in the particle-hole channel.
Examples could be the antiferromagnetic, s- and d- charge density wave
or stripe order parameters.

Many experiments such as angle-resolved photoemission\cite{Damascelli} 
or tunneling in break-junctions\cite{Wilde,Renner,Miyakawa,Zasadzinski} 
suggests that there is only one energy
scale related with the gap and that this scale increases monotonically
with decreasing doping. Recent intrinisc c-axis tunneling spectra in several 
cuprates\cite{Suzuki1,Yurgens1,Krasnov1,Suzuki2,Krasnov2,Krasnov3,Yurgens2,Krasnov4} seem to modify this picture. 
Optimally doped or underdoped samples show at low temperatures two
peaks for positive or negative voltages. The peak at larger voltages 
stays essentially at the same position, but becomes
broader with increasing temperature. With decreasing doping it
moves towards larger voltages. Though this peak  behaves similar to the
one seen in tunneling in break-junctions, it has recently been
argued that heating effects could seriously affect this peak
\cite{Zavaritsky,Yurgens}. The peak
at smaller voltages moves towards zero voltage 
with increasing temperature, looses hereby intensity and vanishes at
T$_c$. Heating effects should be unimportant for the behavior of this
peak. Intrinisc tunnel spectra of this kind have been found both for
double layer and as well as single layer materials\cite{Yurgens2}.
On the other hand, strongly overdoped samples
show only one sharp peak with properties as expected from BCS 
theory\cite{Suzuki1}. 

It is tempting to associate the two peaks observed in the optimally doped
and the underdoped region
with the SC and the pseudogap, as has been done in some of the above
references. In the following we will investigate whether the widely
accepted $t-J$ model supports such a picture. To this end we will
present calculations for the conductance based on a
t-J model
where the two spin components have been generalized to N components
and the leading  diagrams at large N are taken into account.
As discussed in detail in Ref.\cite{Cappelluti} the  phase diagram 
in this
limit has a quantum critical point (QCP) separating at T=0 the normal phase 
at large dopings from a d-CDW state at lower dopings if superconductivity
is omitted. Allowing also for superconductivity the QCP separates
a pure superconducting state from a ground state containing
both superconductivity
and a d-CDW. The properties at optimal doping and in the
underdoped regime are mainly determined by the competition between
superconductivity and the d-CDW. This model thus represents an
example for the above second scenario.
    
\section{Density of states and Fermi surface in the presence of
SC and d-CDW}

The d-CDW order parameter, appropriate for the t-J model at large N, 
is given by 
$\Phi({\bf k}) = {{-i}/{2N_c}}\sum_{{\bf q}\sigma}J({\bf k}-{\bf q})
\langle {\tilde c}^\dagger_{{\bf q}\sigma}\tilde{c}_{{\bf q + Q}\sigma}
\rangle$. $J$ is the Heisenberg coupling, $\tilde{c}^\dagger,\tilde{c}$
are creation and annihilation operators for electrons under the
constraint that double occupancies of lattice sites are excluded, $N_c$
is the number of primitive cells, $\langle ...\rangle$ 
denotes an expectation value, and $\bf Q$ is the wave vector of the d-CDW.
Keeping only the instantaneous term in the 
effective interaction, the superconducting order parameter is 
$\Delta({\bf k}) = {1/{2N_c}} \sum_{\bf q} (J({\bf k}-{\bf q}) -
V_C({\bf k}-{\bf q})) \langle {\tilde c}_{{\bf q}\uparrow} 
{\tilde c}_{-{\bf q}\downarrow}
\rangle$.
As shown in Ref.\cite{Cappelluti} it is in general necessary to include 
the Coulomb potential V$_C$ in order to stabilize the d-CDW with respect 
to phase separation. In the presence
of the two order parameters the operators
$({\tilde c}^\dagger_{{\bf k},\uparrow},\tilde{c}_{-{\bf k},\downarrow},
{\tilde c}^\dagger_{{\bf k+Q},\uparrow},\tilde{c}_{{\bf -k-Q},\downarrow})$
are coupled leading to the following Green's function matrix\cite{Cappelluti}
\begin{equation}
G_0^{-1}(z,{\bf k}) = \left( 
\begin{array}{c c c c}  
z-\epsilon({\bf k}) & -\Delta({\bf k})  & -i\Phi({\bf k})                      &  0                 \nonumber\\
-\Delta({\bf k})    &z+\epsilon({\bf k})
&   0               &i\Phi({\bf \bar{k}})      \\
i\Phi({\bf k})      &   0
&z-\epsilon({\bf\bar{k}})& -\Delta({\bf \bar{k}})  \nonumber\\
          0         &-i\Phi({\bf \bar{k}})
&-\Delta({\bf\bar{k}})&z+\epsilon({\bf \bar{k}})  

\end{array} \right)
\label{matrix}
\end{equation}
$\epsilon({\bf k})$ is the one-particle energy, 
$\epsilon({\bf k}) = -(\delta t +\alpha J)(cos(k_x)+cos(k_y))
-2t'\delta cos(k_x)cos(k_y)  -\mu$,
with $\alpha = 1/N_c \sum_{\bf q} cos({q_x})f(\epsilon({\bf q}))$.
$f$ is the Fermi function, $\delta$ the doping away fom half-filling,
$\mu$ a renormalized chemical potential,
$t$ and $t'$ are nearest and second-nearest
neighbor hopping amplitudes, z a complex frequency, and $\bf {\bar{k}}
= {\bf k-Q}$.

Expressing the expectation values in the order parameters by $G_0$ and
using Eq.(\ref{matrix}) one obtains two coupled equations for the order 
parameters. In the interesting doping region the order parameters
have d-wave symmetry, $\Phi({\bf k})=\Phi \gamma({\bf k}),
\Delta({\bf k})=\Delta \gamma({\bf k})$, with 
$\gamma({\bf k}) =(cos(k_x)-cos(k_y))/2$. The equations for
$\Delta$ and $\Phi$ read then
\begin{eqnarray}
1={{2(J-V_C)}\over N_c} \sum_{\bf k} \sum^4_{\alpha =1}
{ {f(E_\alpha({\bf k}))\gamma^2({\bf k})}\over {\Pi_{\beta \neq \alpha}
(E_\alpha({\bf k})-E_\beta({\bf k}))}} \nonumber \\
\Bigl(\gamma^2({\bf \bar{k}})(\Delta^2+\Phi^2) 
-(E_\alpha({\bf k})
-\epsilon({\bf \bar{k}}))(E_\alpha({\bf k})+\epsilon({\bf \bar{k}})\Bigr),
\label{Ord1}
\end{eqnarray}
\begin{eqnarray}
1={{2J}\over N_c} \sum_{\bf k} \sum^4_{\alpha =1}
{ {f(E_\alpha({\bf k}))\gamma^2({\bf k})}\over {\Pi_{\beta \neq \alpha}
(E_\alpha({\bf k})-E_\beta({\bf k}))}} \nonumber \\
\Bigl(\gamma^2({\bf \bar{k}})(\Delta^2+\Phi^2) 
-(E_\alpha({\bf k})
+\epsilon({\bf k}))(E_\alpha({\bf k})+\epsilon({\bf \bar{k}})\Bigr).
\label{Ord2}
\end{eqnarray}
$V_C$ is the expansion coefficient of the Coulomb potential
in the d-wave channel with the basis function $\gamma$, and
$E_\alpha({\bf k})$ are the four poles of $G_0(z,{\bf k})$ in the
z-plane. At 
\begin{figure}[h]
\centerline{
      \epsfysize=6cm
      \epsfxsize=7cm
      \epsffile{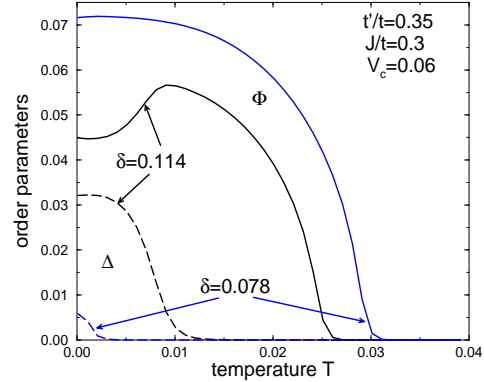}}
\label{fig1}
\caption 
{Order parameters $\Phi$ and $\Delta$
as a function of temperature in units of $t$
for the dopings $\delta = 0.114$ and $0.078$.} 
\end{figure}
\noindent
zero temperature $\Phi$ decreases monotonically 
with increasing $\delta$, whereas $\Delta$ first increases, passes then 
through a maximum at $\delta_0$ and finally decreases again, as shown
in Fig.1 of Ref.\cite{Zeyher1}.
Fig.1 shows the temperature
dependence of $\Phi$ and $\Delta$, calculated with $t'/t=-0.35$
and $J/t=0.3$. The energy unit is $t$.
A repulsive nearest-neighbor Coulomb interaction was also included
with $V_C/t=0.06$. 
According to Fig.1 the temperature dependence of the order
parameters is mean-field like sufficiently away from $\delta_0$. 
Near $\delta_0$ the two order parameters strongly
interact with each other. For instance, for the slightly underdoped case of
$\delta_0=0.114$, the increase of $\Delta$ at low temperatures 
is accompanied by a decrease of $\Phi$ so that the ``total gap''
$\sqrt{\Delta^2+\Phi^2}$ is rather constant at low temperatures.

Fig.2 contains quasi-particle densities for $\delta=0.114$ and $T=0$.
The thin dotted line denotes the density for
vanishing order parameters. It shows a logarithmic divergence
due to the van Hove singularity. The latter lies for the chosen parameters
just below the Fermi energy which corresponds to zero energy. 
The long-dashed curve decribes the case
where the correct finite value for $\Phi$ has been used, but $\Delta$
has been put to zero. This density shows a strongly asymmetric
gap structure with a strong peak on the low and a weaker, splitted
peak at the high-energy side. A closer analysis shows that the lower
peak of this dublett comes from ${\bf k}$-points near the antinodal
points $X$ and $Y$. States near these points are
\vspace{-0.5cm}
\begin{figure}[h]
      \centerline{\hbox{
      \epsfysize=8cm
      \epsfxsize=8cm
      \epsffile{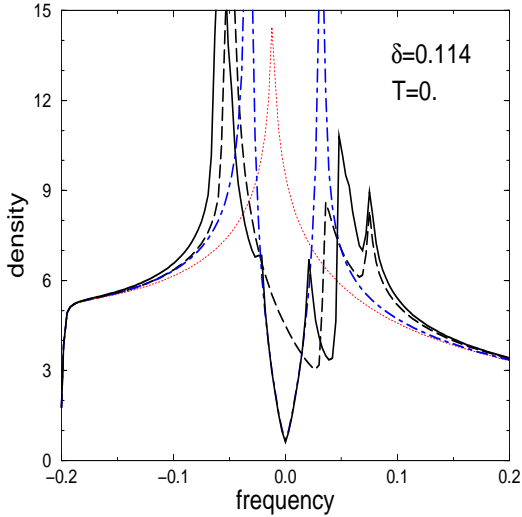}}}
\caption 
{Density of states for $\Phi=\Delta=0$ (thin dotted line), $\Phi \neq 0, 
\Delta=0$ (long-dashed line), $\Phi=0,\Delta \neq 0$ (dash-dotted line),
$\Phi \neq 0, \Delta \neq 0$ (solid line) for $T=0$ and $\delta = 0.114$.}
\label{fig2}
\end{figure}
\noindent
coupled by the
d-CDW and their energies are shifted by the formation of the d-wave gap.
This explains why this peak moves upwards (see the dashed and solid
lines in Fig. 2) in the presence of an additional BCS gap. In contrast
to that, the upper peak of the dublett originates from ${\bf k}$-states
on the boundaries of the reduced Brioulin zone near the points
$(\pi/2-\delta,\pi/2-\delta)$ and equivalent points with $\delta << \pi/2$.
Their energies are mainly determined by the one-particle energies
near this point relative to the Fermi energy and thus less sensitive
to the formation of the gap. This explains why the position of this
higher peak of the dublett is unchanged by the BCS gap, see the dashed and
solid lines in Fig. 2.
The density is everywhere nonzero in the d-CDW state,
in particular, at the Fermi energy. The asymmetry of the density with
respect to zero energy is caused by the asymmetric bare density
due to the proximity of the van Hove singularity. The finite density
of states at the Fermi energy 
allows to lower further the ground state energy by introducing
a superconducting gap.
 The density of states becomes then strictly zero at the
Fermi level and the additional d-wave gap is rather symmetric
with respect ot the Fermi energy. The resulting density of states
for this case $\Phi \neq 0, \Delta \neq 0$ is given by 
the solid line in Fig.2. 
Comparing this line with the dot-dashed line, which 
corresponds to a pure SC state $\Phi = 0, \Delta \neq 0$, one recognizes
that the inner
part of the gap structure looks like a reduced SC gap
with weakly developed edges, at least on the low-energy side.
Fig.3 shows the density of states for $\delta = 0.114$ and
three different temperatures. These temperatures are low enough so
that the main gap edges do not change much because $\sqrt{\Phi^2+\Delta^2}$
is nearly constant. However, the opening of the SC gap in the inner part of
the gap structure can
\begin{figure}[h]
      \centerline{\hbox{
      \epsfysize=8cm
      \epsfxsize=8cm
      \epsffile{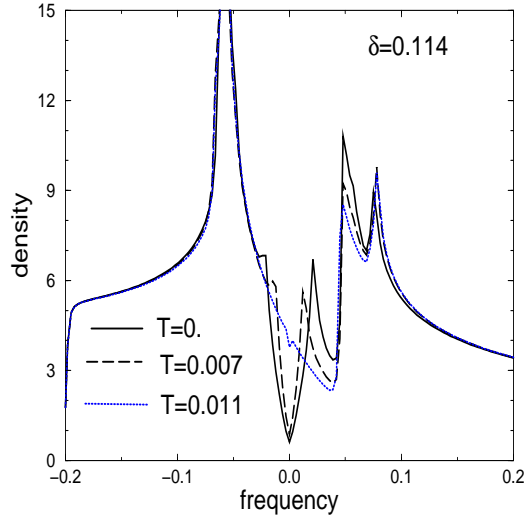}}}
\caption 
{Density of states for $T=0.$ (solid line), $T=0.007$ (dashed line),
$T=0.011$ (dotted line) and $\delta = 0.114$.}
\label{fig3}
\end{figure}
\noindent
clearly be seen as a function of temperature.
\begin{figure}
      \centerline{\hbox{
      \epsfysize=8cm
      \epsfxsize=8cm
      \epsffile{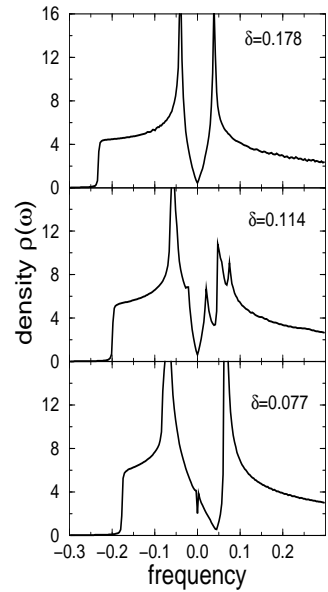}}}
\caption 
{Density of states for three different dopings at $T=0.$} 
\label{fig4}
\end{figure}

Figs.2 and 3 illustrate the fact that the total gap is not 
just one single d-wave gap with an amplitude given by the square root of 
the sum of 
the square of the two gaps. Instead, the SC and CDW gaps interact with
each other, however in such a way that their individual structures can
still be seen in the density of states. This is also apparent in the density
plots for three different dopings in Fig.4. The upper and lower panels
illustrate the difference between the density of states for a SC and a d CDW
gap, respectivley. Some features of the individual gaps 
are still present in the middle panel of Fig.4 which describes the 
case of coexisting SC and d CDW gaps. We find in contrast to Ref.\cite{Kim}
that our self-consistently calculated order parameters yield    
for all considered dopings and hopping parameters densities where the
SC gap lies inside the d-CDW gap.

One important feature of the coexising SC and d-CDW state is that
the two gaps have different locations in
$\bf k$-space: The CDW gap mainly resides near the $X$ and $Y$ points,
and the SC gap near the diagonal. This becomes clear by looking at the
Fermi lines as a function of $\Phi$ in the absence of superconductivity.
Since a finite $\Phi$ implies a doubling of the elementary cell we
have plotted Fermi 
\begin{figure}
      \centerline{\hbox{
      \epsfysize=8cm
      \epsfxsize=8cm
      \epsffile{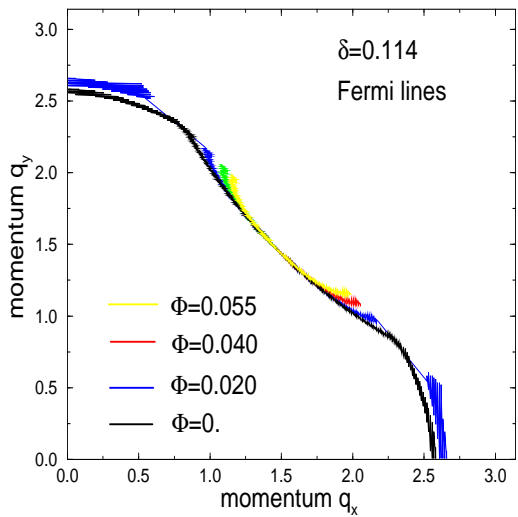}}}
\caption 
{Fermi lines at $T=0$ for four values for $\Phi$ in the absence of
superconductivity.}
\label{fig5}
\end{figure}
\noindent
lines in Fig.5 in the reduced zone scheme, e.g.,
the new Brillouin zone (BZ) is bounded by a straight line between the 
points $X =(\pi,0)$ and
$Y=(0,\pi)$. The black line corresponds to $\Phi = 0$ and describes
the usual normal state Fermi line in the reduced Brillouin zone for the
parameters $t'/t=0.35$ and $J/t=0.3$. For $\Phi = 0.02$ the Fermi line
consists of a long arc around the diagonal ending at the new boundary
of the BZ and two pieces near the $X$ and $Y$ points. This means that the
region around the hot spots becomes first gapped. Increasing $\Phi$ 
only the arc around the diagonal survives and becomes shorter.
The finite density of states at the Fermi energy in the CDW state is due to 
this arc.
Allowing also for SC the arc becomes gapped and the Fermi line
shrinks to one point on the diagonal. The coexistence of SC
and CDW thus becomes possible because the CDW state can lower the
free energy by introducing a SC gap along the arc.

\section{Conductance}

Using lowest-order perturbation theory in the interlayer hopping
the quasi-particle c-axis current J between superconducting layers 
(SIS junction) is given by\cite{Schrieffer}
\begin{eqnarray}
J(V) = {e\over\pi} {1\over N_c^3} \sum_{{\bf k},{\bf q},{\bf k'}}
\ll T_{{\bf k}{\bf q}} T^\ast_{{\bf k'}{\bf q'}}
\gg \int_{-\infty}^\infty d\omega
(f(\omega)-    \nonumber \\ 
f(\omega +eV))   
A_{11}({\bf k'}{\bf k},\omega +eV) 
A_{11}({\bf q}{\bf q'},\omega).
\label{J}
\end{eqnarray}
In Eq.(\ref{J}) $T_{{\bf k}{\bf q}}$ denotes the hopping matrix element
between states with momenta $\bf k$ and $\bf q$ in adjacent layers
and $\ll ... \gg $ an average
over quenched disorder. $f$ is the Fermi function, $V$ the applied voltage,  
and, using the Nambu representation, 
$A_{11}$ the spectral function of the element 11 of the 2x2 
Green's function matrix. The momenta in the above formula refer to
the original (large) BZ. The differential conductance $G(V)$,
which is the main quantity of interest in the following, is defined
as the first derivative of $J$ with respect to $V$.

We make the following Ansatz for the averaged squared tunneling
matrix element
\begin{eqnarray}
\ll T_{{\bf k}{\bf q}} T^\ast_{{\bf k'}{\bf q'}} \gg
= t_\perp^2 \gamma({\bf k}) \gamma({\bf q}) \gamma({\bf k'})
\gamma({\bf q'}) N_c \delta_{{\bf k}-{\bf q},{\bf k'}-{\bf q'}} \nonumber\\
(a \delta_{{\bf k},{\bf q}} +
g({\bf k}-{\bf q})).
\label{average}
\end{eqnarray}
The form factors $\gamma({\bf k})$ determine which electrons in the BZ
are mainly involved in the tunneling process. Results from band structure
calculations\cite{Andersen} suggest the Ansatz\cite{Sandeman} 
\begin{equation}
\gamma({\bf k}) = 1-u +u/2|cosk_x-cosk_y|,
\label{gamma}
\end{equation}
with $0 \leq u \leq 1$. The parameter $u$ interpolates between the
isotropic case $u=0$ and the strongly anisotropic case $u=1$. The latter is
typical for tunneling within a double layer of $CuO_2$ planes, whereas
tunneling between layers in different elementary cells may include also
an isotropic component.   
The first term in the brackets in Eq.(\ref{average}) accounts for
coherent scattering with strength $a$. The second term
in the brackets describes incoherent scattering where $g$ is a smooth
function of the momentum. One may distinguish two cases for
the momentum dependence of $g$. In the case of strong,
localized scatterer $g$ may assumed to be completely independent on momentum. 
If long-ranged random fields are present $g$ is large (small) mainly at small 
(large) momentum transfers. The momentum dependence of $g$ thus can be
modelled by
\begin{equation}
g({\bf k})=g \cdot exp(-|{\bf k}|^2/\Lambda^2), 
\label{g}
\end{equation}
where the
momentum $\Lambda$ interpolates from isotropic to forward scattering, 
described by large and small values for $\Lambda$, respectively.

Since the case of coherent scattering can be obtained from that of
incoherent scattering by replacing $g$ by $4\pi/\Lambda^2$ and taking
the limit $\Lambda \rightarrow 0$
we will first consider incoherent scattering. 
Using Eq.(\ref{average}) and restricting the momenta to the reduced BZ
because of the cell doubling due to the d-CDW we obtain for the tunnel current
\begin{eqnarray}
J(V) &=& {e \over \pi} {t_\perp^2 \over N_c^2} 
{\sum_{{\bf \tilde{k}}{\bf \tilde{q}}}}
\gamma^2({\bf \tilde{k}})
\gamma^2({\bf \tilde{q}}) 
\int^\infty_{-\infty} d\omega (f(\omega)-f(\omega +eV)) \nonumber\\ 
\Bigl( g({\bf \tilde{k}}&-&{\bf \tilde{q}}) 
\sum_{ii'=0,1} A_{1+2i,1+2i'}({\bf \tilde{k}},\omega +eV)
A_{1+2i',1+2i}({\bf \tilde{q}},\omega) \nonumber\\
&+&g({\bf \tilde{k}}-{\bf \tilde{q}}-{\bf Q})(A_{11}({\bf \tilde{k}},
\omega +eV) 
A_{33}({\bf \tilde{q}},\omega) \nonumber\\
&+& A_{33}({\bf \tilde{k}},\omega +eV) A_{11}({\bf \tilde{q}},
\omega))
\Bigr).
\label{JV}
\end{eqnarray}
The tilde on the momenta indicates that these momenta lie in the reduced BZ.
The spectral functions $A_{ij}$ can be assumed to be periodic with resepct
to the reduced BZ. The same is true for the form factors $\gamma$ but not
for the function $g$ which originates from impurity potentials. To make
the expression for $J(V)$ independent of the choice for the reduced BZ
we also translate back the momentum appearing in $g$ to the reduced BZ.
Eq.(\ref{JV}) then becomes
\begin{eqnarray}
J(V)  &=& {e \over \pi} {t_\perp^2 \over N_c^2} 
{\sum_{{\bf \tilde{k}}{\bf \tilde{q}}}}
\gamma^2({\bf \tilde{k}}) \gamma^2({\bf \tilde{q}})
\int^\infty_{-\infty} d\omega (f(\omega)-f(\omega +eV)) \nonumber\\ 
g({\bf \tilde{k}}&-&{\bf \tilde{q}}) 
\Bigl( (A_{11}({\bf \tilde{k}},\omega + eV)+A_{33}({\bf \tilde{k}},\omega+eV))
(A_{11}({\bf \tilde{q}},\omega)         \nonumber\\
&+&A_{33}({\bf \tilde{q}},\omega)) 
+2A_{13}({\bf \tilde{k}},\omega +eV)A_{31}({\bf \tilde{q}},\omega)\Bigr).
\label{JW}
\end{eqnarray}

The first contribution in the big parantheses describes s-wave,
the second one d-wave scattering. Their relative importance is
controlled by the parameter $p_c$ in the function $g$.
If $g$ is independent of the transferred momentum
Eq.(\ref{JV}) simplifies to
\begin{equation}
J(V)  = {{egt_\perp^2} \over \pi} 
\int^\infty_{-\infty} d\omega (f(\omega)-f(\omega +eV)) 
{\tilde{\rho}}(\omega +eV){\tilde{\rho}}(\omega),
\label{JJ}
\end{equation}
with the weighted density
\begin{equation}
{\tilde{\rho}}(\omega) = {1\over N_c} 
{\sum_{\bf \tilde{k}}} \gamma^2({\bf \tilde{k}})
(A_{11}({\bf \tilde{k}},\omega)+A_{33}({\bf \tilde{k}},\omega)).
\label{rho}
\end{equation}

For a SIN junction one usually assumes that $g$ is independent of
momentum. Its current is then obtained from Eq.(\ref{JJ}) by 
identifying one of the two densities with that of the normal metal
$\tilde{\rho}_M$ which can be assumed to be constant. The
resulting conductance of a SIN junction becomes then at not too
high temperatures
\begin{equation}
G_{SIN}(V) = {{eg\tilde{\rho}_M}\over{\pi}}\tilde{\rho}(eV).
\label{SIN}
\end{equation}
Since we will mainly consider SIS junctions in the following conductance
will always refer to SIS junctions unless it is 
stated otherwise.

The simplest case of incoherent scattering corresponds to $u=0$ and 
$\Lambda =\infty$, i.e., where the averaged tunneling matrix element is
independent of all momenta. Using the densities of Fig.4
the resulting conductance curves are 
shown in Fig.6.
In the pure superconducting state (upper panel) the conductance shows 
a broad peak near the gap $2\Delta$ which decays rapidly towards larger
but rather slow towards smaller energies. The conductance is  
positive for all frequencies, especially also above $2\Delta$, which is 
intimately connected to the presence of the  
large and rather constant density of states outside of the gap
region. In the d-CDW case (lower panel in Fig.6) the conductance
curve has two peaks. The higher and dominant one is due to the CDW
gap $2\Phi$. In contrast to the superconductor the d-CDW state has
(neglecting the tiny BCS-gap at the doping $\delta=0.077$) a finite density
of
\vspace{0.cm} 
\begin{figure}[h]
      \centerline{
      \epsfysize=8cm
      \epsfxsize=14cm
      \epsffile{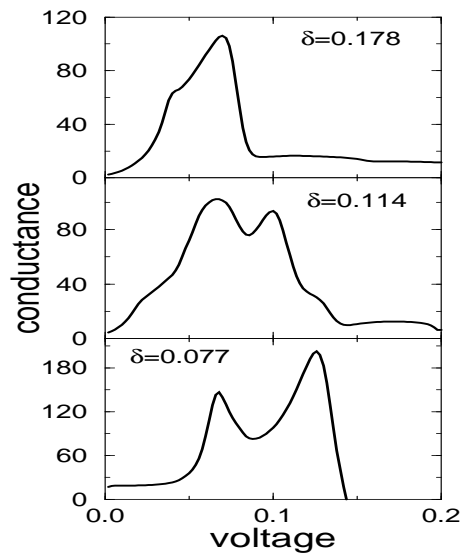}}
\caption 
{Incoherent c-axis conductance calculated for u=0, $\Lambda =\infty$, 
T=0, J=0.3, t'=-0.35, and three different dopings.}
\label{fig6}
\end{figure}
\noindent
states at and near the Fermi energy along the arcs. The folding
of these states with one of the CDW edges causes the lower peak
at about the energy $\Phi$. Well above $2\Phi$ the curve is again rather
constant and partly slightly negative. In the weakly underdoped case
(middle panel in Fig.6) the conductance curve shows essentially two peaks.
They arise due to the folding of the large CDW shoulder at negative
frequencies with the BCS peak and the two splitted CDW shoulders at
positive frequencies, respectively. The BCS gap itself is seen only as a broad
and weak structure at low energies. Some of the features in Fig.6
agree with the tunneling experiments, e.g., the monotonic increase
of the dominant high-frequency peak with decreasing doping and the
appearance of more than one peak in the optimally and underdoped cases.
However, several details of these curves are not found in the experiments:
The peak in the overdoped case is much too
broad compared to that in the intrinisic c-axis tunneling spectra of 
Ref.\cite{Suzuki1}, the lower peak in the underdoped cases is caused by the 
relaxation of electronic states around the CDW gap to 
states near the arcs or the nearby BCS gap and thus does not 
approach zero at $T_c$ as in intrinsic tunneling spectra.
Similar conclusions are reached, following Eq.(\ref{SIN}),
by comparing the experimental 
conductance curves of $SIN$ junctions\cite{Renner,Miyakawa}
with the densities of Fig.4. The monotonic increase of the distance between
the two main peaks with decreasing doping occurs in both cases but the
experiment does not show the asymetry of the theoretical conductance
curve in the underdoped case as well as the additional structures obtained
in the region of coexisting SC and d CDW. 

Things change substantially if the form factor $\gamma({\bf k})$
with a non-zero value for $u$ is taken into account. 
Assuming still a momentum-independent 
function $g$ the tunnel current is now to be calculated from 
the weighted density $\tilde{\rho}$ as given by Eqs.(\ref{JJ}) and 
(\ref{rho}). Fig.7 shows $\tilde{\rho}(\omega)$ for 
\vspace{-0.3cm} 
\begin{figure}[h]
      \centerline{
      \epsfysize=9cm
      \epsfxsize=12cm
      \epsffile{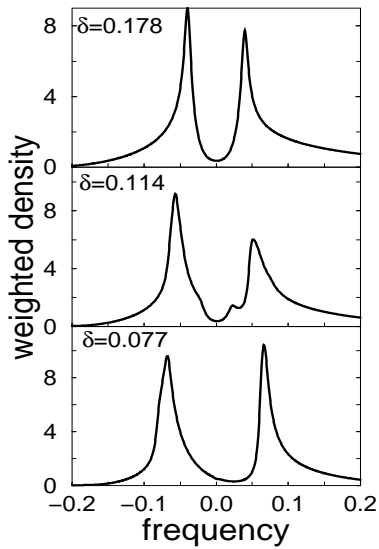}}
\vspace{-0.2cm}
\caption 
{Weighted density $\tilde{\rho}$ as a function of frequency for 
T=0, J=0.3, $\eta=0.004$, t'=-0.35, and three
different dopings $\delta$.}
\label{fig7}
\end{figure}
\vspace{-0.3cm}
\noindent
the extreme anisotropic case
$u=1$ for three different dopings. Most of the background 
contribution to the density has been removed. In the over-
and under-doped cases (upper and lower panels in Fig. 7)
$\tilde{\rho}$ consists of just two rather symmetric
peaks with respect to $\omega = 0$ which are related to the
superconducting and d-CDW gaps, respectively. In the slightly
underdoped regime, where SC and d-CDW coexist, $\tilde{\rho}$
is still rather symmetric with respect to $\omega = 0$ and
consists of four peaks. The peaks at large energies are dominant,
their frequencies are roughly given by $\pm 2 \sqrt{{\Phi}^2+{\Delta}^2}$,
i.e., they describe the ``total'' gap of the two components.
The two weaker and less pronounced peaks at smaller frequencies
are related to the SC gap, which can be concluded from 
their temperature dependence and magnitude of their energies.
The low intensities of these peaks can also be easily understood:
The BCS gap resides on the arcs near the diagonal. The form
factor $\gamma$ with $u=1$ suppresses heavily the tunneling of states   
in this region. It is also interesting to note that only the lower peak
of the splitted high-energy d-CDW edge in Fig.4 for $\delta=0.114$
survives in the corresponding weighted density in Fig.7. This can
easily be understood by noting that the lower (higher) peak of the
dublett is due to ${\bf k}$-states near the antinodal (nodal) points
and thus unaffected (suppressed) by the form factor.

The curves for $\tilde{\rho}$ look in many respects similar to the
experimental SIN conductance curves. In both cases, the spectra are 
dominated by two pronounced peaks lying rather symmetrically with
respect to $\omega = 0$ and whose separation increases monotonically 
with decreasing doping. These peaks evolve in $\tilde{\rho}$ 
very smoothly from a SC to a d CDW state passing through a region where 
both order parameters coexist.
The agreement can be further improved
if one introduces a phenomenological damping in the theoretical curves.
The peaks are then broadened  
\vspace{0.3cm} 
\begin{figure}[h]
      \centerline{
      \epsfysize=8cm
      \epsfxsize=12cm
      \epsffile{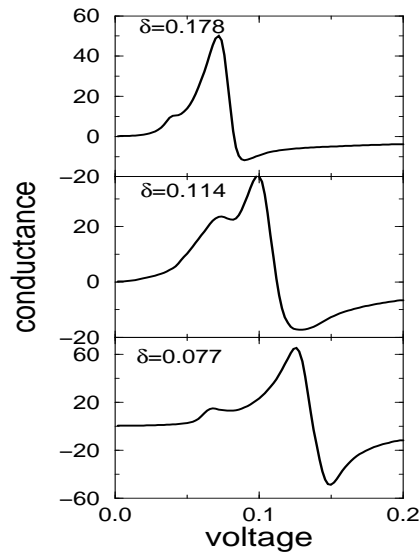}}
\caption 
{Incoherent c-axis conductance calculated for u=1, $\Lambda =\infty$, 
T=0, t'=-0.35, and 
three different dopings.}
\label{fig8}
\end{figure}
\noindent
and the low-energy structures
for $\delta = 0.114$ become invisible. The dip on the high-energy sides of 
the main peaks in the experimental spectra is, however, missing in 
$\tilde{\rho}$ indicating the presence of self-energy effects beyond
a constant damping. 

  Performing the frequency integral in Eq.(\ref{JJ}) with the  
weighted densities $\tilde{\rho}$ one obtains the curves of Fig.8.
The main effect of the inclusion of the anisotropic
form factors $\gamma$ with $u=1$ is the suppression of 
the small quasiparticle excitations near the nodal regions.
This means in the overdoped case $\delta =0.178$ that
the slowly decaying tail of the main peak towards lower voltages
seen in the upper panel of Fig.6  
is substantially suppressed making the peak much sharper. The dip
above $2\Delta$ is also more pronounced than in Fig.6 and the conductance
assumes (small) negative values over a wide region towards higher
voltages. The reason for this negative resistance becomes clear from
a comparison of Figs.4 and 7. Most of the rather constant background
density in Fig.4 has been removed by the anisotropic form factor. However,
just this constant background density is responsible for a positive
and structureless conductance outside of the gap region. 
Similar considerations apply to the underdoped case $\delta=0.077$.
The anisotropic form factor sharpens up somewhat the high voltage peak
and suppresses the lower peak at around $\Phi$ because the states near the 
arcs can no longer contribute much. At the same time the conductance shows
a well-pronounced dip above $2\Phi$ with large negative values
due to the eliminated background density of states. Similar statements
hold for the slightly underdoped case. Here the lower peak at around 
$\Delta + \Phi$, which was in Fig.
6 still the strongest one, is 
suppressed but still visible.   
 
The above calculations show that an incoherent tunneling model
with a momentum independent function $g$ is not able to produce
a peak in the coexistence region which moves towards zero voltage 
if $T$ approaches $T_c$. 
\vspace{-0.3cm} 
\begin{figure}[h]
      \centerline{
      \epsfysize=8cm
      \epsfxsize=12cm
      \epsffile{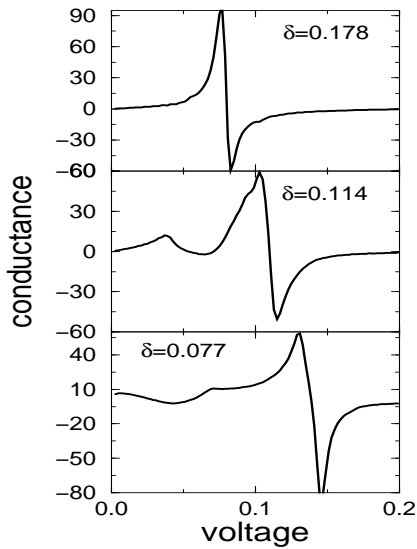}}
\caption 
{Incoherent c-axis conductance calculated for isotropic $u=0$ form factors, a 
momentum cutoff $\Lambda=\pi/8$, $T=0$, $J=0.3$, $t'=-0.35$ and 
three different dopings $\delta$.}
\label{fig9}
\end{figure}
\noindent
We therefore have also studied finite values for $\Lambda$  
in the Gaussian in Eq.(\ref{g}). Experimental evidence for strong
forward scattering in the averaged squared tunneling matrix element
has recently been found from the temperature dependence of the
c-axis penetration depth in $YBa_2Cu_3O_{^+x}$\cite{Hosseini}.
Using the isotropic form factor
$u=0$ and $\Lambda =\pi/8$ Fig.9
shows the conductance for three different dopings at T=0. The
spectra are dominated by a peak at approximately the
frequencies $2\sqrt{\Phi^2+\Delta^2}$. This peak
reflects the doping dependence of the ``total'' gap which increases
monotonically with decreasing doping. In the upper panel the
gap describes a superconducting gap, in the lower panel 
a d-CDW gap and in the middle panel a combination of both.
The doping dependence of the main peak in Fig.9 agrees
well with experimental SIS spectra, see, for instance, Figs. 1 and 2
in\cite{Zasadzinski}, though the dips above the main line are
more pronounced than in the experiment. Also negative conductances are 
only very rarely observed experimentally. 
In the slightly underdoped regime (middle panel in Fig.9) the
two order parameters coexist. The conductance shows in this
case besides of the dominating high-frequency peaks a 
peak near the superconducting part of the gap.
This peak moves towards smaller frequencies
with increasing temperature and vanishes at $T_c$. The exact energy
position of this peak is somewhat below the superconducting part of 
the gap, $2\Delta$. This can easily be understood from Figs.2 and 3:
Due to the interaction between the two gaps part of the BCS shoulder
has been removed by the d-CDW so that the superconducting part of the gap
appears smaller than the canonical value for $2\Delta$. This reduced gap
\vspace{-0.3cm} 
\begin{figure}[h]
      \centerline{
      \epsfysize=8cm
      \epsfxsize=8cm
      \epsffile{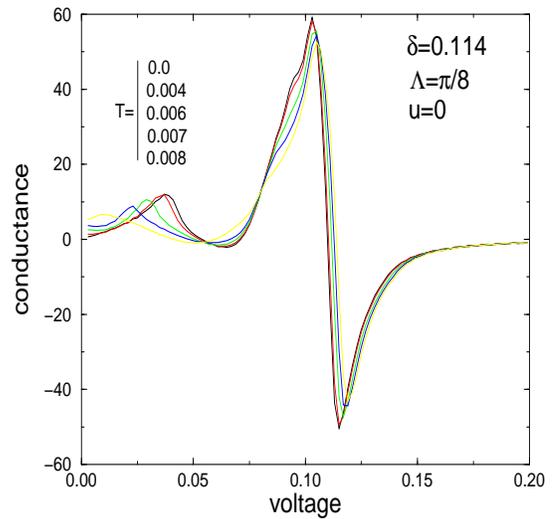}}
\caption 
{Incoherent c-axis conductance for five temperatures using
a Gaussian distributed hopping matrix element with width $\Lambda =\pi/8$.
Increasing temperatures correspond to curves with energetically decreasing 
low-energy peaks.} 
\label{fig10}
\end{figure}
\noindent
exhibits the expected temperature dependence as can be seen from
Fig.3. The upper and lower panels in Fig.9 also show weak
structures at low voltages. A closer inspection, however, reveals that
these structures are caused by the underlying band structure and are
unrelated to the BCS gap.  
The spectra in Fig.9 exhibit well-pronounced
dips at energies somewhat above the main peaks.
These dips, which are also seen in tunneling spectra from break junctions,
have been associated with self-energy effects due to the coupling to some 
boson\cite{Wilde,Zasadzinski}. 
We would like to stress that no self-energy effects
have been taken into account in calculating Fig.9. Simple model calculations
indicate that one finds easily a dip in SIS spectra if the size of the 
region above 
$2\Delta$ where spectral weight piles up because of the formation of the gap
is comparable or smaller than the gap.

Fig.10 shows the temperature dependence of the incoherent conductance for
$\delta = 0.114$ and $\Lambda =\pi/8$.  
The dominating higher peak is  
practically temperature independent for the temperatures shown in the
figure. On the other hand, Fig.1 indicates that both order parameters
vary in the considered temperature interval. One concludes from this
that the higher peak reflects the total gap which is rather independent
of temperature. In contrast to that, the lower peak depends strongly on
temperature. It moves towards zero frequency with increasing
temperature, looses spectral weight and vanishes with vanishing
$\Delta$. Intrinis c-axis tunneling spectra in various cuprates
show essentially the same features as in Fig. 10. In particular,
the observed low-frequency peak, which 
\vspace{-0.3cm} 
\begin{figure}[h]
      \centerline{
      \epsfysize=8cm
      \epsfxsize=14cm
      \epsffile{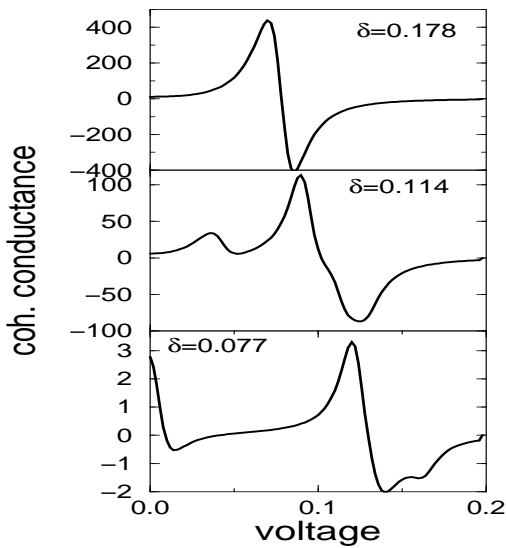}}
\caption 
{Coherent c-axis conductance for anisotropic u=1 form factors, T=0,
J=0.3, t'=-0.35, $\eta=0.01$, and three different dopings $\delta$.}
\label{fig11}
\end{figure}
\noindent
seems not to be affected much
by heating effects, also moves towards zero frequency with increasing
temperature and vanishes near $T_c$. We would like to point out that
this BCS-like peak can be seen in incoherent scattering only for a rather
isotropic form factor $\gamma$. Otherwise, the tunneling of electrons
near the nodal direction, where the SC gap is located, is too much 
suppressed. Another prerequisite is that the averaged tunneling matrix 
element must be strongly momentum-dependent causing strong forward scattering. 

Besides of incoherent tunneling the quasiparticle current always contains
a contribution $J_{coh}$ due to coherent tunneling, 
originating from the first term in the parantheses in Eq.(5). $J_{coh}$ is
given by Eq.(\ref{J}) with $g$ replaced by $a \delta_{{\bf k},{\bf q}}$. 
The explicit expression for $J_{coh}$ thus becomes
\begin{eqnarray}
J_{coh}(V)  = {{eat_\perp^2} \over \pi} &{1\over N_c}&\sum_{\bf \tilde{k}}
\gamma^4({\bf \tilde{k}})
\int^\infty_{-\infty} d\omega (f(\omega)-f(\omega +eV)) \nonumber\\ 
\sum_{ii'=0,1} A_{1+2i,1+2i'}&(&{\bf \tilde{k}},\omega +eV)
A_{1+2i',1+2i}({\bf \tilde{k}},\omega).
\label{Jcoh}
\end{eqnarray}
If the superconducting order parameter is zero, Eq.(\ref{matrix})
reduces to a 2x2 matrix. Calculating explicitly the spectral functions 
from this matrix and performing the frequency integration in the above
integral one finds then that the sum over $i,i'$ yields zero without any 
further approximation. One thus obtains the important result that 
coherent tunneling is zero in a pure d-CDW.   

Fig.11 shows the coherent conductance for $u=1$, $T=0$, and three
different dopings. The spectral functions were obtained from the
Green's functions using the frequency $\omega + i\eta$ with $\eta = 0.01$.
In the overdoped and slightly underdoped case
the curves are similar to those in Fig.9. In particular,
corresponding curves have negative conductances above the main peak
caused by the restriction to small momentum transfers in the tunneling
process. For $\delta = 0.114$ both curves show besides of the main peak
associated with the total gap a second peak at smaller energies with
BCS properties. However, one should note that Fig.9 was calculated with $u=0$
whereas Fig.11 with $u=1$. It is somewhat surprising that coherent
tunneling shows still the BCS peak though most electrons near the 
nodal direction are prevented from tunneling due to the employed strongly
anisotropic form factor. The dominance of small energy features in 
coherent tunneling also is present in the underdoped case $\delta = 0.077$
where the tiny BCS gap causes a sharp structure at very low energies.
One important feature in Fig.11 is related to the absolute values for the
conductance depicted along the y-axis. The coherent conductance 
drops dramatically with decreasing doping. Going to smaller doping values
one finds that $J_{coh}$ becomes zero if the superconducting order
parameter vanishes in agreement with the above analytic result.
Coherent tunneling is non-zero in case of a pure superconductor as
shown by the upper panel in Fig. 11. However, it is zero for a pure
d-CDW state and the d-CDW gap can only be probed in the presence of 
superconductivity. If the identification of the pseudogap phase with a
d-CDW is correct coherent tunneling should vanish in the pseudogap
phase.

Fig.12 illustrates the dependence of the coherent conductance on temperature
for $\delta = 0.114$. With increasing temperature the position of the
low-energy peak and its intensity decrease and approach zero at around
$T=0.0081$ where the superconducting order parameter vanishes.
 The position of the high-energy peak as well 
\vspace{0.cm} 
\begin{figure}[h]
      \centerline{
      \epsfysize=8cm
      \epsfxsize=7cm
      \epsffile{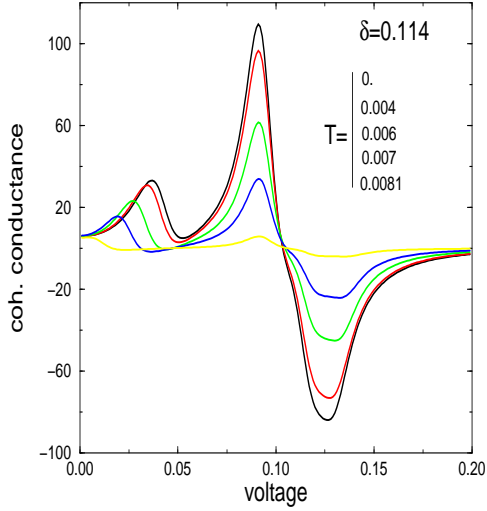}}
\caption 
{Coherent c-axis conductance for anisotropic u=1 form factors, 
J=0.3, t'=-0.35, $\eta=0.01$,  and various temperatures labeled according
to decreasing maxima.}
\label{fig12}
\end{figure}
\noindent
as the dip are rather
independent of temperature but the intensity of the whole high-energy
part drops dramatically with temperature and vanishes at $T=0.0081$.
This again demonstrates that contributions from the d-CDW can only be seen in 
the conductance if the superconducting order parameter is finite, i.e.,
in the coexistence regime.

\section{Self-energy effects}

According to angle-resolved photoemission experiments the generic spectral
function in the superconducting state consists of a well pronounced 
peak followed by a dip and a hump towards larger energies\cite{Damascelli}. 
In concordance with that the electron dispersion shows a kink between 30
and 70 meV below the Fermi energy\cite{Bogdanov}. These features occur 
throughout the
underdoped, optimally doped and the overdoped regime.   
Most of these properties can be reproduced in a model where the
electrons interact with a boson branch (which may be a phonon or a spin
fluctuation)\cite{Eschrig1,Zeyher2,Manske}. 
In the following we assume a dispersionless boson branch 
with a constant dimensionless coupling $\lambda$. In the presence of
SC and d CDW the inverse of the electronic 4x4 Green's function 
$G(z,{\bf k})$ satisfies
\begin{equation}
G^{-1}(z,{\bf k}) = G_0^{-1}(z,{\bf k}) -\Sigma(z,{\bf k}),
\label{G}
\end{equation}
where $G_0^{-1}$ is given by Eq.(\ref{matrix}) and $\Sigma$
is the self-energy. Because the boson-mediated interaction is 
momentum-independent in our model $\Sigma$ has only diagonal elements
and it is $\Sigma_{11}=\Sigma_{22}=\Sigma_{33}
=\Sigma_{44}$, with
\begin{eqnarray}
\Sigma_{11}(z) = \nonumber \\
-{g^2\over N_c}\sum_{\tilde{\bf k}}\Big(
\sum^4_{\alpha=1} {{2\omega_0s(E_\alpha)f(E_\alpha)}\over
{((z-E_\alpha)^2-\omega_0^2)\prod_{\beta \neq \alpha}(E_\alpha -E_\beta)}} 
\nonumber\\
+{{b(-\omega_0)s(z-\omega_0)}\over{{\prod}_\alpha (z-\omega_0-E_\alpha)}}
-{{b(\omega_0)s(z+\omega_0)}\over{{\prod}_\alpha (z+\omega_0-E_\alpha)}}
\Bigr).
\label{Sigma}
\end{eqnarray}
$g^2$ is related to $\lambda$ by $g^2 = \lambda \omega_0/(2N(0))$,
where $N(0)$ is the density of states for one spin direction and
$\omega_0$ the 
frequency of the boson. $E_\alpha$ denote the four 
poles of $G^{(0)}$, $b$ the Bose function, and $N_c$ is two times
the number of allowed momenta $\tilde{\bf k}$. $s(z)$ is given by
\begin{eqnarray}
s(z) = 2z(z^2-(\epsilon^2({\bf k})+\epsilon^2({\bf {k-Q}}))/2
\nonumber \\
-\Delta^2({\bf k}) -\Phi^2({\bf k})).
\label{Zaehler}
\end{eqnarray}
Fig. 13 shows the real and imaginary parts of the retarded
self-energy $\Sigma_{11}(\omega +i\eta)$ at $T=0$. The curves for
different dopings look rather similar. The gap near the Fermi energy
consists of the phonon energy plus the d-wave gap of the SC and/or
d-CDW state. The panels in the figure illustrate the very smooth
transition from a SC to a d-CDW gap with decreasing doping, passing
also very smoothly through the coexistence region of SC and d-CDW.
We used in this and in all the following figures 
\vspace{-0.2cm} 
\begin{figure}[h]
      \centerline{
      \epsfysize=8cm
      \epsfxsize=10cm
      \epsffile{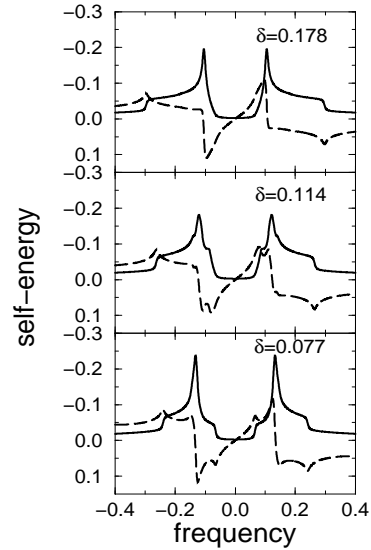}}
\caption 
{Real (dotted line) and imaginary (solid line) parts of the self-energy
for T=0, $\omega_0=0.065$, $\eta=0.004$, and $\lambda=1$.}
\label{fig13}
\end{figure}
\noindent
the value 1 for
the dimensionless coupling constant $\lambda$. This value corresponds
to a change of the slope of the electron dispersion at the kink by a 
factor two in rough agreement with the photoemission data.

Fig.14 shows the SIN conductance using the $u=1$ anistropic hopping form 
factor and including self-energy effects. It is instructive to compare
this figure with the 
\vspace{-0.5cm} 
\begin{figure}[h]
      \centerline{
      \epsfysize=8cm
      \epsfxsize=13cm
      \epsffile{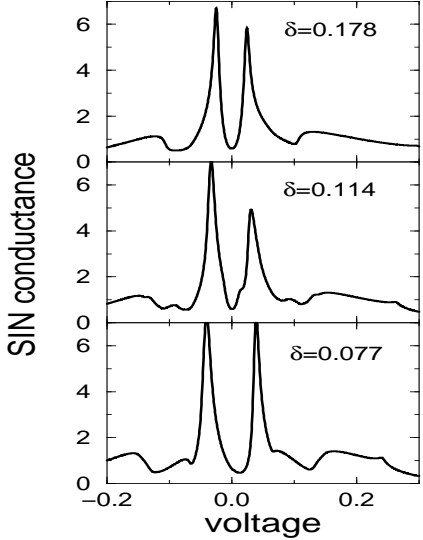}}
\caption 
{SIN conductances with self-energy corrections for T=0, u=1, 
$\omega_0=0.065$, $\eta=0.004$, $\lambda=1$, and three dopings.}
\label{fig14}
\end{figure}
\noindent
analogous Fig.7 where self-energy effects have been
omitted. The BCS-structure seen in Fig.7 for $\delta = 0.114$ has practically
vanished in Fig.14. The peaks in Fig.14 are slightly broader, but the main
effect of the self-energy is to move spectral weight from the main peaks
to the sidebands. In the pure superconducting state at $\delta = 0.178$ 
the sidebands consists of a clear dip and hump whereas in the
two other cases the dip-hump feature is less pronounced. Both dip and hump
move monotonically towards larger voltages with decreasing doping. The
position of the dip in a pure superconductor is approximately half of the 
gap plus
the boson energy. This rule also holds in the d-CDW and the mixed states.
The exact differences between the main peaks and the dip, however,
fluctuate between 0.065 and 0.087 in Fig.14. Though the dip in the SIN 
spectra is solely caused by the interaction with the bosons it may thus 
be 
difficult to determine precisely the boson energy from it.

The solid and dashed lines in Fig.15 are conductance curves for incoherent 
tunneling with and without self-energy effects, respectively. The bosons
do not contribute to the non-diagonal self-energy because of the assumed
momentum-independent coupling to the electrons. As a result,
the bosons diminish both the SC and the d-CDW gaps via
their diagonal self-energies. Consequently, the main peak moves towards
lower frequencies but, considered as a function of doping, this peak
increases monotonically with decreasing doping as in the case without
self-energy. Fig.15 also illustrates that the distance between the dip
and the main peak is similar in the curves with and without self-energy
effects and thus is rather unrelated with the boson energy. 
For instance, at $\delta = 0.178$ the distance between the dip and the
main peak is 0.020 and 0.018 in the case with and without self-energy,
respectively, and thus much smaller than the boson energy of 0.065.
\vspace{-0.3cm} 
\begin{figure}[h]
      \centerline{
      \epsfysize=8cm
      \epsfxsize=12cm
      \epsffile{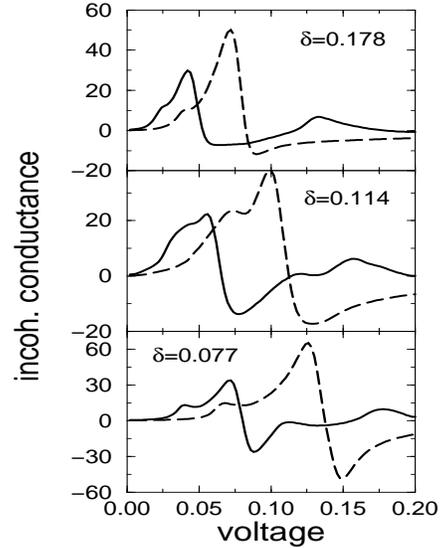}}
\caption 
{Incoherent conductances with (solid line) and without (dashed line)
self-energy corrections for T=0, u=1, $\Lambda =\infty$, 
$\omega_0=0.065$, $\eta=0.004$, $\lambda=1$, and three dopings.}
\label{fig15}
\end{figure}

If the experimental SIS spectra correspond to incoherent tunneling 
the boson energy cannot be obtained from the distance between the main peak
and dip. Since the incoherent SIS spectrum is the folding of the SIN spectrum
in energy the incoherent SIS spectrum could have, in prinicple,
a dip when the lower main peak is multiplied by the dip at positive
voltages and this dip position would be equal to the full gap plus the boson
energy. The upper panel of Fig.15, however, shows that the folding
in energy does not lead to a dip just below the maximum of the sideband. 
The main effect of the self-energy in Fig.15 is to shift
spectral weight from the main peak to the sideband consisting of a
broad hump which monotonically moves towards larger voltages with
decreasing doping. This hump is due to the folding of the (occupied) lower
main peak with the (unoccupied) upper sidebands in Fig.14. 
The solid and dashed lines in Fig.16 represent coherent conductance curves
with and without self-energy effects, respectively. We have omitted 
curves for the strongly underdoped case $\delta=0.077$ because they
are smaller by two order of magnitudes due to the smallness of the
supercondcuting order parameter in this case. Self-energy effects shift
the main peaks to smaller energies, diminish somewhat the
regions of negative conductance, and create weak sidebands.
 Figs.7 and 14-16 suggest that self-energy effects and thus the nature of
the boson spectrum appear more clear-cut in the SIN than in the SIS
spectra. For instance, the dip in the SIN spectrum is soley caused by 
self-energy effects whereas that in the SIS spectrum is  
present even in the absence of any self-energy. The curves for the
purely superconducting case in Figs.14 and 15 are similar to those
published in Refs.\cite{Eschrig1,Eschrig2}.
\vspace{-1.5cm} 
\begin{figure}[h]
      \centerline{
      \epsfysize=10cm
      \epsfxsize=12cm
      \epsffile{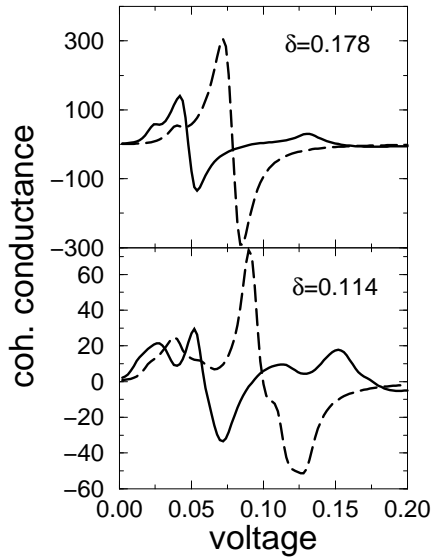}}
\caption 
{Coherent conductances with (solid line) and without (dashed line)
self-energy corrections for T=0, u=1, 
$\omega_0=0.065$, $\eta=0.004$, $\lambda=1$, and two dopings.}
\label{fig16}
\end{figure}

\section{Conclusions}

The $t-J$ model exhibits in the employed large-N limit a d-CDW phase
at lower dopings besides of the superconducting phase which is 
a natural candidate for the pseudogap phase observed in the cuprates.
The density of states in the pure d-CDW state is strongly reduced near the
Fermi level but still everywhere finite. This means that only part
of the Fermi lines of the normal state are destroyed by the d-CDW
and that the remaining Fermi lines form arcs around the nodal direction
ending at the boundaries of the reduced Brillouin zone. With 
decreasing doping the length of the arcs become shorter. The ground
state energy of the d-CDW can be lowered by introducing a d-wave
superconducting gap near the arcs which explains the occurrence of a 
coexistence region of SC and d-CDW. Because the two gaps are well
separated in $\bf k$-space (the d-CDW gap resides near the antinodal,
the superconducting gap near the nodal direction) features of the individual 
gaps survive even in the coexistence regime.   

In order to test the applicability of the above picture to cuprates
we have calculated coherent and incoherent conductances and compared
them with experimental spectra from break-junctions and intrinsic
tunneling spectroscopy. We find good evidence that the tunneling
matrix element between layers is strongly anisotropic, suppressing 
tunneling of electrons near the nodal direction, which is in
agreement with band structure arguments. Incoherent tunneling 
thus probes mainly electrons near the maximal gap at the X and
Y points. This gap transforms in a very smooth way from a
superconducting gap at large dopings to a d-CDW gap at small dopings
passing continuously through the coexistence regime. Calculated
incoherent conductances thus fit best to the observed spectra from
break-junctions which are characterized by one peak moving monotonically
to larger voltages with decreasing doping. We find that coherent tunneling
is only non-zero for a non-vanishing superconducting order parameter. Below
$T_c$ it shows in spite of the anisotropic tunneling matrix element two
peaks which can be associated with SC and d-CDW. The appearance of
a low-energy peak in the calculated coherent tunneling spectrum,
which moves to 0 if $T_c$ is approached 
from below, is unique for intrinsic tunneling spectra. From this we
conclude that tunneling in stacked, intrinsic junctions is dominated by 
coherent tunneling and that the appearance of the low-energy peak
related to superconductivity supports models with two competing order
parameters in the underdoped region. Including self-energy effects due to the
coupling of electrons to a dispersionless boson branch as suggested by
ARPES removes part of the regions of negative resistances and also
creates sidebands which, at least in the case of SIN junctions, resemble
those which have been measured.   

The authors thank Secyt and the BMBF ( Project ARG 99/007) for financial 
support and A. Yurgens and V.M. Krasnov for useful discussions.

\end{multicols}
\end{document}